\documentclass[twocolumn,showpacs,preprintnumbers,amsmath,amssymb]{revtex4}

\usepackage{graphicx}
\usepackage{dcolumn}
\usepackage{bm}

\begin{document}

\preprint{APS/123-QED}

\title{Influence of Striction Effects on the Multicritical Behavior
of Homogeneous Systems}

\author{S.V. Belim}
 \email{belim@univer.omsk.su}
\affiliation{%
Omsk State University, 55-a, pr. Mira, Omsk, Russia, 644077
\textbackslash\textbackslash
}%

\date{\today}

\begin{abstract}
A field-theoretical description of the behavior of homogeneous,
elastically isotropic, compressible systems characterized by two
order parameters at the bicritical and tetracritical points is
presented. For three-dimensional Ising-like systems, a similar
description is performed in the two-loop approximation in three
dimensions. The renormalization group equations are analyzed, and fixed points
corresponding to different types of multicritical behavior are
determined. It is shown that the effect of elastic strains causes
a change from a bicritical behavior to a tetracritical one and
leads to the appearance of a wide variety of multicritical points.
\end{abstract}

\pacs{64.60.-i}
\maketitle

Earlier, it was shown [1] that, because of striction effects,
elastic strains lead to the appearance of multicritical points
that are absent in the phase diagrams of corresponding
incompressible substances.

The subject of this paper is the study of the influence of
striction effects on systems whose phase diagrams already contain
multicritical points of the bicritical or tetracritical type. In the
first case, a multicritical point corresponds to the intersection
of two lines of secondorder phase transitions and one line of
first-order phase transitions, and in the second case, it
corresponds to the intersection of four lines of second-order
phase transitions. In the immediate vicinity of a multicritical
point, the system exhibits a specific critical behavior
characterized by the competition between different types of
ordering. As a result, at a bicritical point, one critical
parameter is displaced by another. A tetracritical point allows
the existence of a mixed phase with the coexistence of different
types of ordering. Such systems [2] can be described by
introducing two order parameters that are transformed according to
two irreducible representations.

In structural phase transitions that occur in the absence of the
piezoelectric effect in the paraphase, elastic strains act as a
secondary order parameter whose fluctuations are not critical in
most cases [3]. Since, in the critical region, the main
contribution to the striction effects comes from the dependence of
the exchange integral on the distance, only elastically isotropic
systems are considered in this paper.

The model Hamiltonian of the system has the form
\begin{eqnarray}\label{gam1}
&&H_0=\int d^Dx\Big[\frac{1}{2}(\tau_1+\nabla^2)\Phi(x)^2
+\frac{1}{2}(\tau_1+\nabla^2)\Psi(x)^2\nonumber\\
&&+\frac{u_{10}}{4!}(\Phi(x)^2)^2
+\frac{u_{20}}{4!}(\Psi(x)^2)^2+\\
&&+\frac{2u_{30}}{4!}(\Phi(x)\Psi(x))^2 +g_1y(x)\Phi(x)^2\nonumber\\
&&+g_2y(x)\Psi(x)^2 +\beta y(x)^2\Big],\nonumber
\end{eqnarray}

Here, $\Phi(x)$ and $\Psi(x)$ are the fluctuating order parameters;
$u_{01}$ and $u_{02}$ are positive constants;
$\tau_1\sim|T-T_{c1}|/T_{c1}$ and $\tau_2\sim|T-T_{c2}|/T_{c2}$,
where $T_{c1}$ and $T_{c2}$ are the phase transition temperatures
for the first- and second-order parameters, respectively;
$y(x)=\sum\limits_{\alpha=1}^3u_{\alpha\alpha}(x)$, where
$u_{\alpha \beta}$ is the strain tensor;
$g_1$ and $g_2$ are the quadratic striction parameters;
$\beta$ is a constant characterizing the elastic properties of the crystal;
and $D$ is the space dimension. In this Hamiltonian, integration with
respect to the components that depend on the nonfluctuating
variables, which do not interact with the order parameters, has
already been already performed.

Changing to the Fourier transforms of the variables in Eq. (1),
one obtains the Hamiltonian of the system in the form
\begin{eqnarray}\label{gam2}
&&H_0=\frac 12\int d^Dq(\tau _1+q^2)\Phi_q\Phi_{-q}\nonumber\\
&&+\frac 12\int d^Dq(\tau _2+q^2)\Psi_q\Psi_{-q}\\
&&+\frac{u_{01}}{4!}\int d^D{q_i}(\Phi_{q1}\Phi_{q2})(\Phi_{q3}\Phi_{-q1-q2-q3})\nonumber\\
&&+\frac{u_{02}}{4!}\int d^D{q_i}(\Psi_{q1}\Psi_{q2})(\Psi_{q3}\Psi_{-q1-q2-q3})\nonumber\\
&&+\frac{2u_{03}}{4!}\int d^D{q_i}(\Phi_{q1}\Phi_{q2})(\Psi_{q3}\Psi_{-q1-q2-q3})\nonumber\\
&&+g_1\int d^Dqy_{q1}\Phi_{q2}\Phi_{-q1-q2}\nonumber\\
&&+g_2\int d^Dqy_{q1}\Psi_{q2}\Psi_{-q1-q2}
+\frac{g^0_1}{\Omega}y_0\int d^Dq\Phi_{q}\Phi_{-q}\nonumber\\
&&+\frac{g^0_2}{\Omega}y_0\int d^Dq\Psi_{q}\Psi_{-q}
+2\beta\int d^Dqy_qy_{-q} +2\frac{\beta_0}{\Omega}y_0^2\nonumber
\end{eqnarray}

Here, the components $y_0$ describing uniform strains are separated.
According to [1], such a separation is necessary, because the
nonuniform strains $y_q$ are responsible for the acoustic phonon
exchange and lead to long-range interactions, which are absent for
uniform strains.

Let us determine the effective Hamiltonian that depends on only
the strongly fluctuating order parameters $\Phi$ and $\Psi$ of the system as
follows:
\begin{eqnarray}\label{3}
\exp \{-H[\Phi,\Psi]\}=B\int \exp \{-H_{0}[\Phi,\Psi,y]\}\prod
dy_q
\end{eqnarray}

If the experiment is performed at constant volume, the quantity $y_0$
is a constant, and the integration in Eq. (3) only goes over the
nonuniform strains, while the uniform strains do not contribute to
the effective Hamiltonian. In an experiment at constant pressure,
the term $P\Omega$. is added to the Hamiltonian, with the volume being
represented in terms of the strain tensor components in the form
\begin{eqnarray}\label{6_1}
\Omega=\Omega_0 [1+\sum\limits_{\alpha =1}u_{\alpha\alpha}+
\sum\limits_{\alpha \neq
\beta}u_{\alpha\alpha}u_{\beta\beta}+O(u^3)]
\end{eqnarray}

and the integration in Eq. (3) also performed over the uniform
strains. According to [4], the inclusion of quadratic terms in Eq.
(4) may be important at high pressures and for crystals with
strong striction effects. As a result, one obtains
\begin{eqnarray}\label{gam3}
&&H =\frac 12\int d^Dq(\tau _1+q^2)\Phi_q\Phi_{-q}\nonumber\\
&&+\frac 12\int d^Dq(\tau _2+q^2)\Psi_q\Psi_{-q}\\
&&+\frac{v_{01}}{4!}\int d^D{q_i}(\Phi_{q1}\Phi_{q2})(\Phi_{q3}\Phi_{-q1-q2-q3})\nonumber\\
&&+\frac{v_{02}}{4!}\int d^D{q_i}(\Psi_{q1}\Psi_{q2})(\Psi_{q3}\Psi_{-q1-q2-q3})\nonumber\\
&&+\frac{2v_{03}}{4!}\int d^D{q_i}(\Phi_{q1}\Phi_{q2})(\Psi_{q3}\Psi_{-q1-q2-q3})\nonumber\\
&&+\frac{z_1^2-w_1^2}{2}\int d^D{q_i}(\Phi_{q1}\Phi_{-q1})(\Phi_{q2}\Phi_{-q2})\nonumber\\
&&+\frac{z_2^2-w_2^2}{2}\int d^D{q_i}(\Psi_{q1}\Psi_{-q1})(\Psi_{q2}\Psi_{-q2})\nonumber\\
&&+(z_1z_2-w_1w_2)\int d^D{q_i}(\Phi_{q1}\Phi_{-q1})(\Psi_{q2}\Psi_{-q2})\nonumber \\
&& v_{01}=u_{01}-12z_1^2, \ \ v_{02}=u_{02}-12z_2^2,\nonumber\\
&& v_{03}=u_{03}-12z_1z_2,\ \  z_1=\frac{g_1}{\sqrt\beta},\nonumber\\
&& z_2=\frac{g_2}{\sqrt\beta},\ \ w_1=\frac{g^0_1}{\sqrt\beta_0},\nonumber\\
&& w_2=\frac{g^0_2}{\sqrt\beta_0}\nonumber
\end{eqnarray}

This Hamiltonian leads to a wide variety of multicritical points.
As for incompressible systems, both tetracritical
$(v_3+12(z_1z_2-w_1w_2))^2<(v_1+12(z_1^2-w_1^2))(v_2+12(z_2^2-w_2^2))$
and bicritical
$(v_3+12(z_1z_2-w_1w_2))^2\geq(v_1+12(z_1^2-w_1^2))(v_2+12(z_2^2-w_2^2))$
behaviors are possible. In addition, the striction effects may
give rise to multicritical points of higher orders.

In the framework of the field-theoretical approach [5], the
asymptotic critical behavior and the structure of the phase
diagram in the fluctuation region are determined by the
Callan–Symanzik renormalization group equation for the vertex
parts of the irreducible Green's functions. To calculate the $\beta$
and $\gamma$ functions as functions involved in the Callan–Symanzik equation
for renormalized interaction vertices $u_1, u_2, u_3, g_1, g_2, g_1^{(0)}, g_2^{(0)}$
or complex vertices $z_1$, $z_2$,  $w_1$, $w_2$, $v_1$, $v_2$, $v_3$, which are more
convenient for the determination of the multicritical behavior, a
standard method based on the Feynman diagram technique and on the
renormalization procedure was used [6]. As a result, the following
expressions were obtained for the â functions in the two-loop
approximation:
\begin{eqnarray}\label{8}
&&\beta_{v1}=-v_1+\frac{n+8}{6}v_1^2+\frac{m}{6}v_3^2\nonumber\\
&&-\frac{41n+190}{243}v_1^3-\frac{23m}{243}v_1v_3^2-\frac{2m}{27}v_3^3,\nonumber\\
&&\beta_{v2}=-v_2+\frac{m+8}{6}v_2^2+\frac{n}{6}v_3^2-\frac{41m+190}{243}v_2^3\nonumber\\
&&-\frac{23n}{243}v_2v_3^2-\frac{2n}{27}v_3^3,\\
&&\beta_{v3}=-v_1+\frac{2}{3}v_3^2+\frac{(n+2)}{6}v_1v_3+\frac{m+2}{6}v_2v_3\nonumber\\
&&-\frac{5(n+m)+72}{486}v_3^3-\frac{23(n+2)}{486}v_1^2v_3-\nonumber\\
&&-\frac{23(m+2)}{486}v_2^2v_3-\frac{n+2}{9}v_1v_3^2-\frac{m+2}{9}v_2v_3^2,\nonumber\\
&&\beta_{z1}=-z_1+\frac{n+2}{3}v_1z_1+2nz_1^3+2mz_1z_2^2+\frac{m}{3}v_3z_2\nonumber\\
&&-\frac{23(n+2)}{243}v_1^2z_1-\frac{7m}{243}v_3^2z_1-\frac{2m}{27}v_3^2z_2,\nonumber\\
&&\beta_{z2}=-z_2+\frac{m+2}{3}v_2z_2+2mz_2^3+2nz_1^2z_2+\frac{n}{3}v_3z_1\nonumber\\
&&-\frac{23(m+2)}{243}v_2^2z_2-\frac{7n}{243}v_3^2z_2-\frac{2n}{27}v_3^2z_1,\nonumber\\
&&\beta_{w1}=-w_1+\frac{n+2}{3}v_1w_1+4nz_1^2w_1-2mw_1^3\nonumber\\
&&+4mz_1z_2w_2-2mw_1w_2^2
+\frac{m}{3}v_3w_2-\nonumber\\
&&-\frac{23(n+2)}{243}v_1^2w_1-\frac{7m}{243}v_3^2w_1-\frac{2m}{27}v_3^2w_2,\nonumber\\
&&\beta_{w2}=-w_2+\frac{m+2}{3}v_2w_2+4mz_2^2w_2-2nw_2^3\nonumber\\
&&+4nz_1z_2w_1-2nw_1^2w_2+\frac{n}{3}v_3w_1-\nonumber\\
&&-\frac{23(m+2)}{243}v_2^2w_2-\frac{7n}{243}v_3^2w_2-\frac{2n}{27}v_3^2w_1.\nonumber
\end{eqnarray}

It is well known that the perturbative series expansions are
asymptotic and the vertices of the interactions of the order
parameter fluctuations in the fluctuation region are sufficiently
large for Eqs. (6) to be directly applied. Therefore, to extract
the necessary physical information from the expressions derived
above, the Pade-Borel method generalized to the multiparameter
case was used. The corresponding direct and inverse Borel
transformations have the form
\begin{eqnarray}
&& f(v1,v2,v3,z1,z2,w1,w2)=\nonumber\\
&&=\sum\limits_{i_1,...,i_7}
c_{i_1...i_7}v_1^{i_1}v_2^{i_2}v_3^{i_3}
z_1^{i_4}z_2^{i_5}w_1^{i_6}w_2^{i_7}\nonumber\\
&&=\int\limits_{0}^{\infty}e^{-t}F(v_1t,v_2t,v_3t,z_1t,z_2t,w_1t,w_2t)dt,  \\
&& F(v1,v2,v3,z1,z2,w1,w2)=\nonumber\\
&&=\sum\limits_{i_1,...,i_7}
\frac{\displaystyle c_{i_1,...,i_7}}
{\displaystyle(i_1+...+i_7)!}v_1^{i_1}v_2^{i_2}v_3^{i_3}
z_1^{i_4}z_2^{i_5}w_1^{i_6}w_2^{i_7}.
\end{eqnarray}

For the analytic continuation of the Borel transform of
a function, a series in an auxiliary variable $\theta$ is introduced:
\begin{eqnarray}
&&\tilde{F}(v_1,v_2,v_3,z_1,z_2,w_1,w_2,\theta)=\\
&&=\sum\limits_{k=0}^{\infty}\theta^k\sum\limits_{i_1,...,i_7}
\frac{\displaystyle c_{i_1...i_7}}{\displaystyle k!}v_1^{i_1}v_2^{i_2}v_3^{i_3}
z_1^{i_4}z_2^{i_5}w_1^{i_6}w_2^{i_7}\delta_{i_1+...+i_7,k},\nonumber
\end{eqnarray}

and the [L/M] Pade approximation is applied to this
series at the point $\theta=1$. This approach was proposed
and tested in [7] for describing the critical behavior of
systems characterized by several vertices of interaction
of the order-parameter fluctuations. The property of the
system retaining its symmetry when using the Pade
approximants in variable $\theta$ is essential for the description
of multivertex models.

In the two-loop approximation, the â functions were
calculated using the [2/1] approximant. The character
of critical behavior is determined by the existence of a
stable fixed point satisfying the set of equations
\begin{equation}
\beta_{i}(v_1^*,v_2^*,v_3^*,z_1^*,z_2^*,w_1^*,w_2^*)=0\ \ \   (i=1,...,7).
\end{equation}

The requirement that the fixed point be stable is
reduced to the condition that the eigenvalues $b_i$  of the
matrix
\begin{eqnarray}  \displaystyle
&&B_{i,j}=\frac{\partial\beta_i(v_1^*,v_2^*,v_3^*,z_1^*,z_2^*,w_1^*,w_2^*)}{\partial{v_j}}\\
&&(v_i,v_j \equiv v_1^*,v_2^*,v_3^*,z_1^*,z_2^*,w_1^*,w_2^*)\nonumber
\end{eqnarray}
lie in the right complex half-plane.

The resulting set of resummed $\beta$ functions contains
a wide variety of fixed points lying in the physical
region of the vertex values with vi $v_i\geq 0$.

A complete analysis of the fixed points, each of
them corresponding to the critical behavior of a single
order parameter, was presented in our recent publication
[8]. Now, I consider the combined critical behavior
of two order parameters.

The analysis of the values and stability of the fixed
points offers a number of conclusions. The bicritical
fixed point of an incompressible system ($v_1 =
0.934982, v_2 = 0.934982, v_3 = 0.934982, z_1 = 0, z_2 = 0,
w_1 = 0, w_2 = 0$) is unstable under the effect of uniform
strains ($b_1 = 0.090, b_2 = 0.523, b_3 = 0.667, b_4 = 0.521,
b_5 = 0.002, b_6 = 0.521, b_7 = 0.002$). The striction effects
lead to the stabilization of the tetracritical fixed point of
a compressible system ($v_1 = 0.934982, v_2 = 0.934982,
v_3 = 0.934982, z_1 = 0, z_2 = 0, w_1 = 0, w_2 = 0, b_1 = 0.090,
b_2 = 0.523, b_3 = 0.667, b_4 = 2.144, b_5 = 0.267, b_6 =
5.223, b_7 = 0.882$).

The stability of other multicritical points cannot be
investigated in terms of the described model, because
the calculations lead to a degenerate set of equations.
The degeneracy is eliminated by considering the
Hamiltonian with allowance for the terms of higher
orders in both the strain tensor components and the
fluctuating order parameters.

Thus, the striction-caused interaction of the fluctuating
order parameters with elastic strains leads to the
transition from the bicritical behavior to the tetracritical
one and also to the appearance of new multicritical
points with their own types of critical behavior in the
phase diagram of the substance.

\def\baselinestretch{1.0}

\end{document}